\documentclass{article}
\usepackage{PRIMEarxiv}
\usepackage[numbers]{natbib}
\let\printorcid\relax 
\usepackage[flushleft]{threeparttable}
\usepackage{microtype}
\usepackage{graphicx}
\usepackage{caption}
\usepackage{xcolor}
\usepackage{amsmath,amssymb,amsfonts}
\usepackage{algorithmic}
\usepackage{textcomp}
\usepackage{hyperref}       
\usepackage{url}            
\usepackage{enumitem}       
\usepackage{booktabs}       
\usepackage{nicefrac}       
\usepackage{lipsum}
\usepackage{tabularx} 
\usepackage{multirow}
\usepackage{verbatim} 
\usepackage[nameinlink,capitalize]{cleveref}

\usepackage{fancyhdr}       

\title{Wavelet-Based~Multi-Class~Seizure~Type~Classification~System
}

\author{
  Hezam Albaqami, Ghulam~Mubashar~Hassan, Amitava Datta\\
  Department  of  Computer  Science  and  Software  Engineering,   \\
  The  University  of  Western  Australia \\
  Perth\\
  \texttt{hezam.albaqami@research.uwa.edu.au} \\
    \texttt{18-Jun-2021}\\
}

\begin{document}

\let\WriteBookmarks\relax
\def\floatpagepagefraction{1}
\def\textpagefraction{.001}

\maketitle

\begin{abstract}
Epilepsy is one of the most common brain diseases that affect more than 1\% of the world's population. It is characterized by recurrent seizures, which come in different types and are treated differently. Electroencephalography (EEG) is commonly used in medical services to diagnose seizures and their types. The accurate identification of seizures helps to provide optimal treatment and accurate information to the patient. However, the manual diagnostic procedures of epileptic seizures are laborious and highly-specialized. Moreover, EEG manual evaluation is a process known to have a low inter-rater agreement among experts. This paper presents a novel automatic technique that involves extraction of specific features from EEG signals using Dual-tree Complex Wavelet Transform (DTCWT) and classifying them. We evaluated the proposed technique on TUH EEG Seizure Corpus (TUSZ) ver.1.5.2 dataset and compared the performance with existing state-of-the-art techniques using overall F1-score due to class imbalance seizure types. Our proposed technique achieved the best results of weighted F1-score of 99.1\% and 74.7\% for seizure-wise and patient-wise classification respectively, thereby setting new benchmark results for this dataset.
\end{abstract}

\section{Introduction}
\label{INTRO}
\subsection{Background}

Epilepsy is the most widespread brain disease among children and adults after stroke \cite{Subasi2019afterstroke}. It is defined as "a sudden and recurrent brain malfunction and is a disease that reflects an excessive and hypersynchronous activity of the neurons within the brain"~\cite{sanei2013eeg}. Over 60 million of the world population are diagnosed with epilepsy, whose defining feature is recurrent seizures. Such seizure attacks impair the brain’s normal functions, leading the patient to be vulnerable and unsafe. 

Seizures are medically classified into two main categories: focal seizure or generalized seizure, depending on the extent to which regions of the brain are affected. Focal seizures are seizures that originate and affect a circumscribed region of the brain. Focal seizures are further classified into simple or complex, based on the patient’s level of awareness. Generalized seizures, on the other hand, involve most areas of the brain. Based on motor and non-motor symptoms, generalized seizure classifications can be absence, tonic, atonic, clonic, tonic-clonic, or myoclonic seizures\cite{scheffer2017ilae,Fisher2017}. Classification of seizure is very essential for accurate diagnosis and treatment.

Identifying the type of seizure, although sometimes difficult, can be done by clinical observation, referencing medical history and demographic information, and is supported by general brain imaging techniques such as EEG, Magnetoencephalography (MEG), and fMRI~\cite{alotaiby2017epileptic,goldenberg2010overview}. EEG is the most practical and cost-effective tool to diagnose epilepsy nowadays~\cite{Obeid2018}. Video-EEG monitoring is often required to support the decision for seizure classification~\cite{Liu2020}. 

For treatment, seizures can be controlled in most cases (up to 70\%) of patients by consuming medication to achieve a steady-state concentration in the blood. Surgical intervention is another option for certain conditions. For up to 20\% of epileptic patients, there is no medical treatment that exists to control seizures~\cite{sanei2013eeg}. The accurate identification of the type of seizures influences medication choice and provides information to patients, families, researchers, and clinicians caring for patients with epilepsy~\cite{Fisher2017, Roy2019}.

It is a challenging task to classify the type of seizure accurately. Several factors make the classification difficult. Firstly, some types of seizures share the same clinical and EEG symptoms. For instance, it has been shown that even for a highly experienced neurologist, sometimes it is hard to distinguish between focal and generalized seizures~\cite{panayiotopoulos2005optimal}. Secondly, in some cases, it is required to perform long-term monitoring (i.e., video-EEG monitoring), which may last for days~\cite{Obeid2018}. Therefore, manual analysis of these long recordings require a substantial amount of effort and time from neurologists. 

In addition, signal interpretation is known to have a low-inter-rater agreement which fully depends on the level of expertise of the expert. Moreover, inter-subject variability significantly adds to the difficulties associated with diagnosis of an epileptic seizure, leading to a variety of manifestations of the same type of seizures across different patients, and sometimes for the same individual over time. Finally, signal artifacts also hinder the correct interpretation of EEG. With these challenges, in a field that already has shortage of healthcare experts, computer-aided diagnostic (CAD) methods have great potential to support decision-making in the diagnosis of such a critical disease.

\subsection{Review of related work}
A considerable amount of research has been published on automated seizure detection and prediction. However, the automatic classification of seizure types received little attention due to two main reasons. Firstly, the difficulties inherent in the classification problem for seizure types, and secondly lack of clinical data~\cite{Asif}.

Since the start of this century, considerable research outcomes have focused on the automation of epileptic seizure diagnoses~\cite{Liu2020,Roy2019}. Generally, the procedure of automatic seizure analysis involves two phases: feature extraction and classification~\cite{SHOEIBI2021113788}. Various methods have been proposed for feature extraction over time, including time-domain~\cite{altunay2010epileptic}, frequency-domain~\cite{polat2007classification} and time-frequency domain~\cite{subasi2007eeg}. 

Recently time-frequency methods became popular due to inclusion for both time and frequency features. Among time-frequency methods, Wavelet Transforms (WT) based feature extraction is the most promising method to extract robust features from EEG signals~\cite{FAUST201556}. The strategies in wavelet-based feature extraction from EEGs use Continues Wavelet Transform (CWT)\cite{ur2013automatedCWT}, Discrete Wavelet Transform (DWT)\cite{subasi2019comparison}, Wavelet Packet Decomposition (WPD)~\cite{albaqami2021automatic,subasi2019comparison}, Tunable-Q Factor Wavelet Transform (TQWT)\cite{HASSAN2016247} and Dual tree wavelet transform (DTCWT)~\cite{swami2016novel}. 

Regarding the availability of clinical data, it has been observed that during last few years hospitals and universities have made appreciative efforts to encourage research on automatic diagnosis of epileptic seizures by generating large volumes of openly available clinical EEG data. One of the most extensive publicly obtainable EEG data, the Temple University Hospital EEG Corpus (TUH EEG), is comprised of 14.000 subjects and has more than 25,000 clinical recordings~\cite{obeidEEGCORPUSS}. The Corpus has various subsets; each focusing on different scopes of research interests. The TUH EEG Seizure Corpus (TUSZ)~\cite{dataset2018}, one of the subsets, is created to motivate research on developing high-performance epileptic seizure detection algorithms using advances in machine learning algorithms~\cite{dataset2018}. This dataset is manually annotated of seizure events based on archival neurologists’ report and careful examinations of the signals by students and neurologists from Temple University~\cite{dataset2018}. 
The seizure events in the TUSZ are labeled with eight different types of seizures: Focal Non-specific Seizure (FNSZ), Generalized Non-specific Seizure (GNSZ), Simple Partial Seizure (SPSZ), Complex Partial Seizure (CPSZ), Absence Seizure (ABSZ), Tonic Seizure (TNSZ), Tonic-clonic Seizure (TCSZ) and Mayoclinc Seizure (MYSZ). The details of these labels are presented in \cref{table:seiz_definition.0_1}. The corpus team continuously updates the corpus and \cref{table:data_stats_v1.4.0} presents the distribution of data for the last two versions of TUSZ.

\begin{table}
\scriptsize
\centering
\caption{Seizure Types descriptions for TUH EEG Seizure Corpus (TUSZ)}
\label{table:seiz_definition.0_1}
 \begin{tabularx}{0.9\columnwidth}{cX} 
 \toprule
                Seizure Type &   Seizure Description \\\midrule
                FNSZ & Focal seizures which cannot be specified with its type.\\
                SPSZ & Partial seizures during consciousness which is specified by clinical signs only.\\
                CPSZ & Partial Seizures during unconsciousness which is specified by clinical signs only.\\
                GNSZ & Generalized seizures which cannot be further specified with its type.\\
                ABSZ & Absence Discharges observed on EEG where patient loses consciousness for few seconds (also known as \textit{Petit Mal}).\\
                TNSZ & Stiffening of body during seizure (EEG effects disappears).\\
                TCSZ & At first stiffening and then jerking of body (also known as \textit{Grand Mal}).\\
                MYSZ & Myoclonous jerks of limbs.\\\bottomrule
\end{tabularx}
\end{table}

To the best of our knowledge, we found only eight published research studies which used TUSZ for the problem of seizure type classification, the summary is presented in~\cref{table:seizure-wise-all-studies}. Regarding the seven (7) types of seizure classification, Roy et al.~\cite{Roy2019} applied Extreme Gradient Boosting (XGBoost) and KNN to classify the EEG signals into seven classes of seizures. The study reported F1-scores of 85.1\% and 90.1\% for XGBoost and \textit{K}-Nearest Neighbor (KNN) respectively. Similarly, Aristizabal et al.~\cite{AhmedtAristizabal2020} developed a deep learning model known as Neural Memory Networks (NMN) to classify seven types of seizures. The study reported 94.50\% F1-score. In another study related to seven-class problem, Asif et al. \cite{Asif} applied a deep learning framework, called as SeizureNet, with ensemble learning and multiple DensNets that achieved the results of 95\% F1-score. 

Raghu et al.~\cite{raghu2020eeg} extracted EEG image features using a pretrained Google’s Inception 3, and classified them using Support Vector Machine (SVM), achieving an accuracy of 88.3\% to classify between 7 types of seizure classes and normal class. Similarly, in~\cite{sriraam2019convolutional}, a convolutional neural network (CNN) model \textit{AlexNet} is applied to classify EEG images based on the technique of short-time Fourier Transform (STFT) to classify 7 types of seizure and non-seizures class. The study yielded an accuracy of 84.06\%. Liu et al. \cite{Liu2020} applied a hybrid bilinear model consisting of CNN and Long Short-Term Memory (LSTM) to classify between 8-types of seizures. The study reported  97.4\% F1-score. 

For 4-class classification of seizure, Wijayanto et al.~\cite{Wijayanto2019} applied empirical mode decomposition (EMD) to EEGs for feature extraction and quadratic SVM for classification. The study reported an accuracy of 95\%. In another study, Ramadhani et al.~\cite{Ramadhani2019} applied EMD, Mel Frequency Cepstral Coefficients (MFCC) and Independent Component Analysis (ICA) to EEG data for feature extraction and SVM for classification between 4-classes of seizures, and achieved 91.4\% accuracy. For 3-classes of seizure classification, Saric et al.~\cite{Saric2020} developed a Field Programmable Gate Array (FPGA) based framework for the classification of generalized and focal epileptic seizures using a feed-forward multi-layer neural network and achieved an accuracy of 95.14\%.

In spite of good performance reported in aforementioned studies, it is expected that the reported techniques cannot be used in real world situations as the studies either did not report the performance when tested on data from new patients or reported lower performance. Out of the eight studies presented in \cref{table:seizure-wise-all-studies}, only two studies \cite{Roy2019,Asif} considered the generalization of their proposed techniques. Both studies mentioned a considerable decrease in the performance of their system where the performance decreased by 45\%. This shows that there is still a large gap for advancement for better generalization capability for the classification systems.

\begin{table*}
\centering
\caption{Data distribution for different types of seizure in two versions of TUSZ.}
\label{table:data_stats_v1.4.0}
 \begin{tabular}{lccccccc} 
 \toprule
             &   \multicolumn{2}{c}{{No. of seizure events}} &   \multicolumn{2}{c}{{Duration (Seconds)}}  &   \multicolumn{2}{c}{{No. of patients}}\\
    \cmidrule{2-7}
    {Seizure Type}&ver.1.4.0&ver.1.5.2 &ver.1.4.0&ver.1.5.2 &ver.1.4.0&ver.1.5.2 \\
    \midrule
    FNSZ & 992& 1836& 73466& 121139 & 109&150	\\
    GNSZ & 415&	583& 34348&59717 &44&81	\\
    CPSZ & 342& 367& 33088&36321& 34&41	\\
    ABSZ & 99&	99& 852&852& 13&12	\\
    TNSZ & 67&	62& 1271&1204& 2&3 \\
    TCSZ & 50&	48& 5630&5548 &11&12	\\
    SPSZ & 44&	52& 1534& 2146&2&3 \\
    MYSZ & 3&	3& 1312&1312& 2&2  \\ \bottomrule
\end{tabular}
\end{table*}
It is interesting to observe from~\cref{table:seizure-wise-all-studies} that the authors of these studies chose different number of seizure classes, ranging from a three-class problem to an eight-class problem for seizure type classification. The reason behind the choice of the number of classes is not discussed in most of these studies. The authors of \cite{Roy2019,Asif,AhmedtAristizabal2020,raghu2020eeg,sriraam2019convolutional} excluded the seizure type MYSZ from their experiments because the signals of this type were only recorded from two patients (see \cref{table:data_stats_v1.4.0}). However in~\cite{Liu2020}, authors chose to utilize all seizure types in the dataset regardless of the number of patients. \cref{table:seizure-wise-all-studies} presents the investigated seizure types for each study.

It can be observed from \cref{table:seiz_definition.0_1} that in TUSZ, there are six specific types of seizure and two non-specific general types. From a pathological point of view, these types are not completely disjoint but form a hierarchal sub-grouping~\cite{Fisher2017,AhmedtAristizabal2020}. It has been stated in~\cite{Obeid2018} that when there is inadequate evidence to label the type of seizure confidently; the corpus team tends to label an event as either focal non-specific or generalized non-specific based on the seizure’s focality and locality~\cite{dataset2018}. Both of these types are not medically distinct from one another, whereas, SPNS and CPSZ are more specific types of FNSZ, and ABSZ, TNSZ, TNSZ and MYSZ are more specific types of GNSZ~\cite{Fisher2017,AhmedtAristizabal2020}. Thus, considering the label FNSZ as a unique type of seizure against the specific focal types CPSZ and SPSZ might cause the classifier not to perform well, and similarly for the classification of GNSZ. 

Therefore, in this study, we are considering two different classification problems. In the first problem, each label is considered in the dataset as a unique seizure type and results are compared with the existing state-of-the-art results. Whereas, the second problem is the introduction of a new challenge which is more important pathologically that deals with the specific seizure type classification to investigate the effect of the non-specific labels in TUSZ (5-class classification).

\begin{table*}[!ht]
\scriptsize
 \caption{Summary of existing state-of-the-art techniques for seizure classification}
  \centering
  
  \begin{threeparttable}
  \begin{tabularx}{0.9\linewidth}{XlXll}
  
    \toprule
    Method     & No. of seizure classes & Classes considered & Features  & Performance (\%)\\
    \midrule
     Transfer learning Inceptionv3\cite{raghu2020eeg} & 8\tnote{*} &GNSZ, FNSZ, SPSZ, CPSZ, ABSZ, TNSZ, TCSZ, NORM\tnote{+}   & SFFT\     &    88.3 Accuracy    \\ 
     AlexNet\cite{sriraam2019convolutional}   & 8\tnote{*} &GNSZ, FNSZ, SPSZ, CPSZ, ABSZ, TNSZ, TCSZ, NORM\tnote{+}   &     SFFT\  &84.06 Accuracy    \\
     
      CNN+LSTM+MLP\cite{Liu2020}     & 8& GNSZ, FNSZ, SPSZ, CPSZ, ABSZ, TNSZ, TCSZ, MYSZ         & SFFT\    & 97.40 F1-score \\
      
    SeizureNet Ensemble CNNs\cite{Asif}     & 7& GNSZ, FNSZ, SPSZ, CPSZ, ABSZ, TNSZ, TCSZ         &   FFT &95 F1-score \\
    Plastic NMN\cite{AhmedtAristizabal2020}& 7 &GNSZ, FNSZ, SPSZ, CPSZ, ABSZ, TNSZ, TCSZ    & FFT     &    94.5 F1-score   \\
    K-NN\cite{Roy2019}     & 7& GNSZ, FNSZ, SPSZ, CPSZ, ABSZ, TNSZ, TCSZ        &   FFT &90.1 F1  \\
    XGBoost\cite{Roy2019}     & 7& GNSZ, FNSZ, SPSZ, CPSZ, ABSZ, TNSZ, TCSZ        &   FFT &85.1 F1-score  \\
   
    SVM~\cite{Ramadhani2019}     & 4\tnote{*} &GNSZ, FNSZ, TCSZ, NORM\tnote{+} & MFCC+HD+ICA &   91.4 Accuracy  \\
    
    FPGA-based ANN\cite{Saric2020}     &3\tnote{*}&GNSZ, FNSZ, NORM\tnote{+} & CWT      &    95.14 Accuracy \\
    SVM\cite{Wijayanto2019}     & 4 &GNSZ, FNSZ, SPSZ, TNSZ  & EMD    &    95 Accuracy  \\
     
    \bottomrule
  \end{tabularx}
  \begin{tablenotes}                                      
    \item[*Including non-seizure EEG class.] \item[+Normal EEGs.] 
    \end{tablenotes}
    \end{threeparttable}
  \label{table:seizure-wise-all-studies}
\end{table*}

In order to solve the above mentioned problems, we propose a novel technique that focuses on wavelet-based machine learning methods for automatic seizure type classification in multi-channel EEG recordings. We only utilized EEG data and decomposed the EEG signals into different levels of components using DTCWT to extract specific features from these decomposed components. We used shift-invariant DTCWT for feature extraction from biomedical signal and its classification which is done for the first time in literature for seizure type classification. Moreover, we tested our proposed technique on the largest available seizure EEG database, containing various types of epileptic seizures. In order to ensure the effectiveness and generalization of our technique, we thoroughly tested our proposed technique across subjects in addition to normal testing. The experimental results show that our proposed novel technique performs well for both problems of seizure-type classification.

The rest of this paper is organized as follows: \cref{Method} discusses information about the data utilized in this research and the details of our proposed technique. \cref{RESULTS} presents the evaluation methodology and the analyses on the obtained results. A thorough discussion is provided in \cref{discussion}. \cref{conclusion} concludes the article with future research plan.

\section{Methods}\label{Method}
\subsection{Data} 
\textcolor{black}{
Our study is based on TUSZ ver. 1.5.2 dataset~\cite{dataset2018}, which is the largest publicly available dataset released in 2020. This dataset includes 3,050 seizure events, consisting of various seizure morphologies and recorded from over 300 different patients. The TUSZ is obtained from historical hospital data from Temple University Hospital (TUH), where clinical EEG data was retrieved and stored in .EDF format. The signals were recorded based on 10/20 system. \cref{table:data_stats_v1.4.0} presents the details of distribution of TUSZ. The EEG signals in TUSZ are annotated based on electrographic, electro-clinical and clinical manifestations. More details about the dataset can be found in~\cite{dataset2018} and the dataset is available online at corpus website\footnote{\url{https://isip.piconepress.com/projects/tuh_eeg/}}. The seizure type MYSZ is excluded from our study due to its scarcity in the dataset, as it is recorded from just two patients in the recently released version. As mentioned earlier this decision is in accordance with previous research studies in the same field [9], [11], [24]–[26].}

\begin{table}
\caption{EEG Channel names included in our study} 
\centering 
\begin{tabular}{cc|||cc} 
\toprule 

    \#&Channels&\#&Channels \\ 
\hline 
    1&FP1-F7&2&F7-T3\\3&T3-T5&4&T5-O1\\5&FP2-F8&6&F8-T4\\7&T4-T6&8&T6-O2\\9&T3-C3&10&C3-CZ\\
    11&CZ-C4&12&C4-T4\\13&FP1-F3&14&F3-C3\\15&C3-P3&16&P3-O1\\17&FP2-F4&18&F4-C4\\19&C4-P4&20&P4-O2\\

\bottomrule 
\end{tabular}
\label{table:TCP} 
\end{table}

\subsection{Proposed technique}
\begin{figure*}[!h] 
    \centering
    \includegraphics[width=0.9\linewidth]{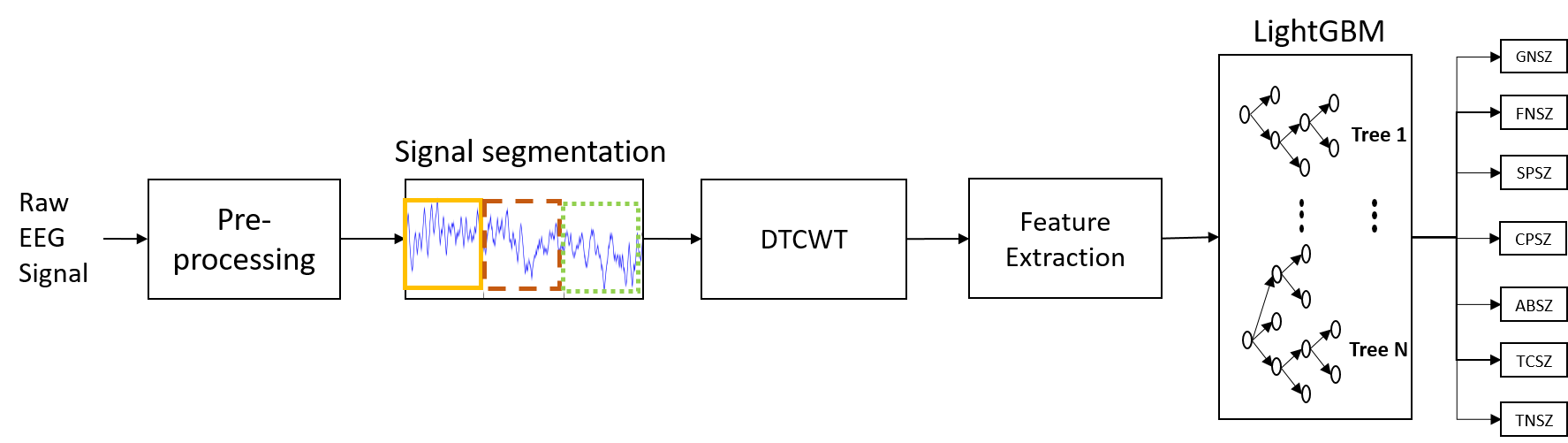}
    \caption{Overall architecture of the proposed technique.}
    \label{fig:overall_archi}
\end{figure*}

Our proposed technique involves multiple steps which include, preprocessing of the data, extracting the important features and then classifying them. The architecture of our proposed technique is presented in \Cref{fig:overall_archi} and all the steps are explained below.

\subsubsection{Preprocessing}

\textcolor{black}{The TUSZ is collected from archival hospital data at Temple University Hospital and the clinical EEG data was recovered from CD ROMs and saved in EDF format~\cite{dataset2018}. All the data do not have the same montage and sampling rate. As a result, we performed some initial procedures to generalize the input data prior to feature extraction. Firstly, the EEG segments which are exclusively responsible for seizures were extracted from the dataset. This was achieved using the annotated file provided in the dataset, including the start and the stop time of each seizure event. We have excluded the seizure type MYSZ due to its scarcity in the dataset. After extracting} \textcolor{black}{the seizure events, we used the Transverse Central Parietal (TCP) montage to accentuate spikes activity~\cite{Roy2019}. Montage is a differential view of the data, which consists of differencing the signals collected from two electrodes (e.g., Fp1-F7, F7-T3)~\cite{ref5}. In fact, neurologists are very particular about the type of montage used when interpreting an EEG~\cite{ref5,ferrell2020temple}. Temple University Hospital (TUH) also reported the same and mentioned that it helps in noise reduction by differences signals~\cite{ferrell2020temple,ref5}. Different experiments on montage selection has been done in~\cite{ref5, shah2017optimizing} and TCP was found to be the most efficient montage that helps different machine learning algorithms to detect seizures. Secondly, we re-sampled all recordings at 250Hz. Finally, we cropped each extracted signal into equally non-overlapped segments such that each segment is of the length of two seconds.  This choice was influenced by~\cite{Roy2019} where the authors have investigated different window length and they reported that the two second window length of the signal is the most optimal choice to achieve the best classification results. In summary, we took the following preprocessing steps in sequence to generalise the input data for processing:}

\begin{enumerate}
    \item  Used the transverse central parietal (TCP) montage to accentuate spikes activity. \cref{table:TCP} presents the EEG channels considered in our study.
    \item Re-sampled all recordings at 250Hz.

    \item Cropped the signal into equally non-overlapped segments such that each segment is of two seconds, resulting in 500 data points.
\end{enumerate}

After the initial preprocessing steps, the input data was generalised and ready to processed for transformation.

\subsubsection{Feature extraction}
Wavelet Transform (WT) methods have been employed successfully to solve various non-stationary signal problems~\cite{FAUST201556,tuncer2020surface,georgieva2019wavelet}, including EEG~\cite{albaqami2021automatic}. WT is a spectral estimation method that provides another representation of the signal at different scale components. Discrete Wavelet Transform (DWT) is one of the WT's most popular technique that decomposes a given signal $x[k]$ into a mutually orthogonal set of wavelets through convolution with filter banks. For $j$ levels of decomposition, a signal $x[k]$ is passed through two bandpass filters: high $h[.]$ and low $l[.]$ starting from $j=1$. The output of each level are two down sampled components: Approximation $Aj$ and Detail $Dj$ which are represented as:

\begin{equation}
D_{j}[i]=\sum_{k} x[k] \cdot h[2 \cdot i-k]
\end{equation}

\begin{equation}
A_{j}[i]=\sum_{k} x[k] \cdot l[2 \cdot i-k]
\end{equation}

\begin{figure}[!h] 
    \centering
    \includegraphics[width=0.8\linewidth]{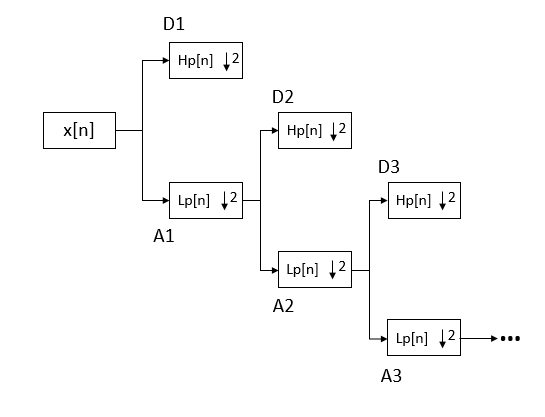}
    \caption{The structure of three-scale level  Discrete Wavelet Transform (DWT).}
    \label{fig:DWT}

\end{figure}

The approximation component $Aj$ can be further decomposed into another level of $A_{j+1}$ and $D_{j+1}$ as shown in \Cref{fig:DWT} until the maximum or required level of $j$ is reached.

\begin{figure}[!h] 
    \centering
    \includegraphics[width=0.8\linewidth]{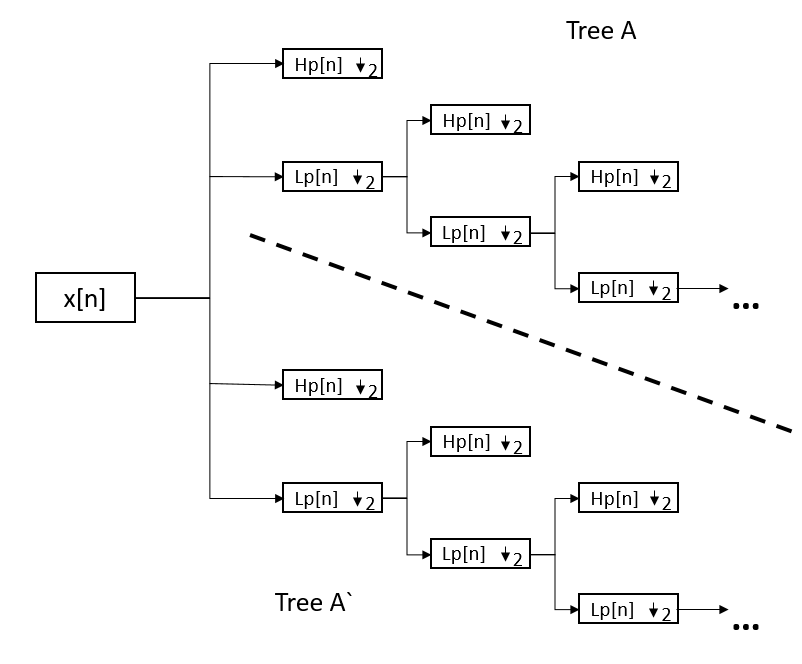}
    \caption{The structure of three scale level of DTCWT.}
    \label{fig:DTCWT}
\end{figure}
\textcolor{black}{DWT has many successful applications however has some drawbacks, such as insufficient information in high-frequency components,} \textcolor{black}{shift-variance, low directionality and absence of phase shift. Over time, different enhancement have been introduced to cover the shortcomings of DWT. Dual-tree Complex Wavelet Transform (DTCWT) is an extension of DWT which was proposed by Kingsbury~\cite{KINGSBURY2001234} and developed later by Selesnick et al.~\cite{selesnick2005dual}. It uses extra double low-pass filters and another two high-pass filters to produce four components at each level which include real and imaginary parts. DTCWT can be imagined as two parallel DWTs as shown in \Cref{fig:DTCWT}.  This transformation is approximate shift-invariant and directionally selective in two and higher dimensions, which are very important in applications such as pattern recognition and signal analysis. Therefore, DTCWT has less shift variance and more directionality as compared to DWT.}

\textcolor{black}{In our proposed technique, we decomposed the EEG signals into four levels using DTCWT using the Python library DTCWT \cite{wareham2014dtcwt}\footnote{\url{https://github.com/rjw57/dtcwt/tree/0.12.0}}. The parameters used for the decomposition were set manually based on trial and error after experimenting with different parameters of decomposition \textcolor {black}{levels} based on the empirical performance and computational efficiency. The decomposition process produces real and imaginary parts of complex wavelet coefficients and we selected the magnitude of the complex coefficients. After the decomposition, we computed a set of statistical features from each of the coefficients. The computed features and their corresponding mathematical representations are presented below. For mathematical representations, $M$ is the length of the signal in each sub-band which is taken as 500 in this study while $Y\{y_1,y_2, ....y_M\}$ and $Z\{z_1,z_2, ....z_M\}$ are two adjacent sub-bands \cite{kevric2017comparison}.}

\begin{enumerate}\label{stfeatures}
\item Mean absolute values (MAV) of the coefficients in each sub-band, $\mu$ .
\begin{equation}
\mu=\frac{1}{M} \sum_{j=1}^{M}\left|y_{j}\right|
\end{equation}
\item Average power (AVP) of the coefficients in each sub-band, $\lambda$. 
\begin{equation}
\lambda=\sqrt{\frac{1}{M} \sum_{j=1}^{M} y_{j}^{2}}
\end{equation}
\item Standard deviation (SD) of the coefficients in each sub-band, $\sigma$. 
\begin{equation}
\sigma=\sqrt{\frac{1}{M} \sum_{j=1}^{M}\left(y_{j}-\mu\right)^{2}}
\end{equation}
\item Ratio of the absolute mean values (RMAV) of adjacent sub-bands, $\chi$ . 
\begin{equation}
\chi=\frac{\sum_{j=1}^{M}\left|y_{j}\right|}{\sum_{j=1}^{M}\left|z_{j}\right|}
\end{equation}
\item Skewness (skew) of the coefficients in each sub-band, $\phi$. 
\begin{equation}
\phi=\sqrt{\frac{1}{M} \sum_{j=1}^{M} \frac{\left(y_{j}-\mu\right)^{3}}{\sigma^{3}}}
\end{equation}
\item Kurtosis (Kurt) of the coefficients in each sub-band, $\phi_k$ . 
\begin{equation}
\phi_k=\sqrt{\frac{1}{M} \sum_{j=1}^{M} \frac{\left(y_{j}-\mu\right)^{4}}{\sigma^{4}}}
\end{equation}
\end{enumerate}

\textcolor{black}{The features across all the statistical coefficients corresponding to this interval signal are stacked together which forms a 6×5 (statistical feature x DTCWT coefficients) feature matrix. We have 20 channels in TCP montage as mentioned in Table IV. Therefore, our resulting feature matrix is of size 20 x 6 x 5 (No of channels x statistical features x DTCWT coefficients) which is flattened to 1 x 600 vector for classification.}

\subsubsection{Feature analysis}
\begin{figure*}[!h] 
    \centering
    \includegraphics[width=\textwidth]{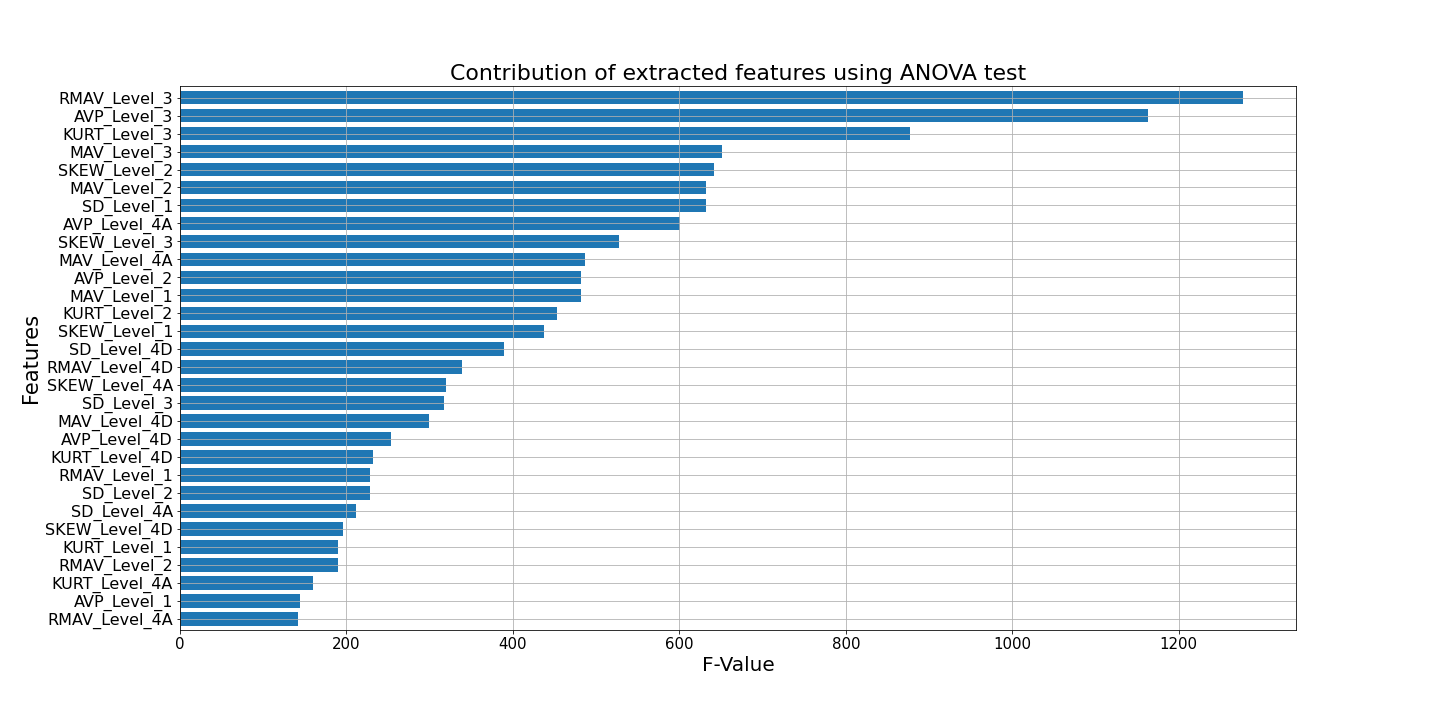}
    \caption{The obtained F-values of the features, using one-way ANOVA test. }
    \label{fig:anova}
\end{figure*}

\begin{figure*}[!h] 
    \centering
    \includegraphics[width=\textwidth]{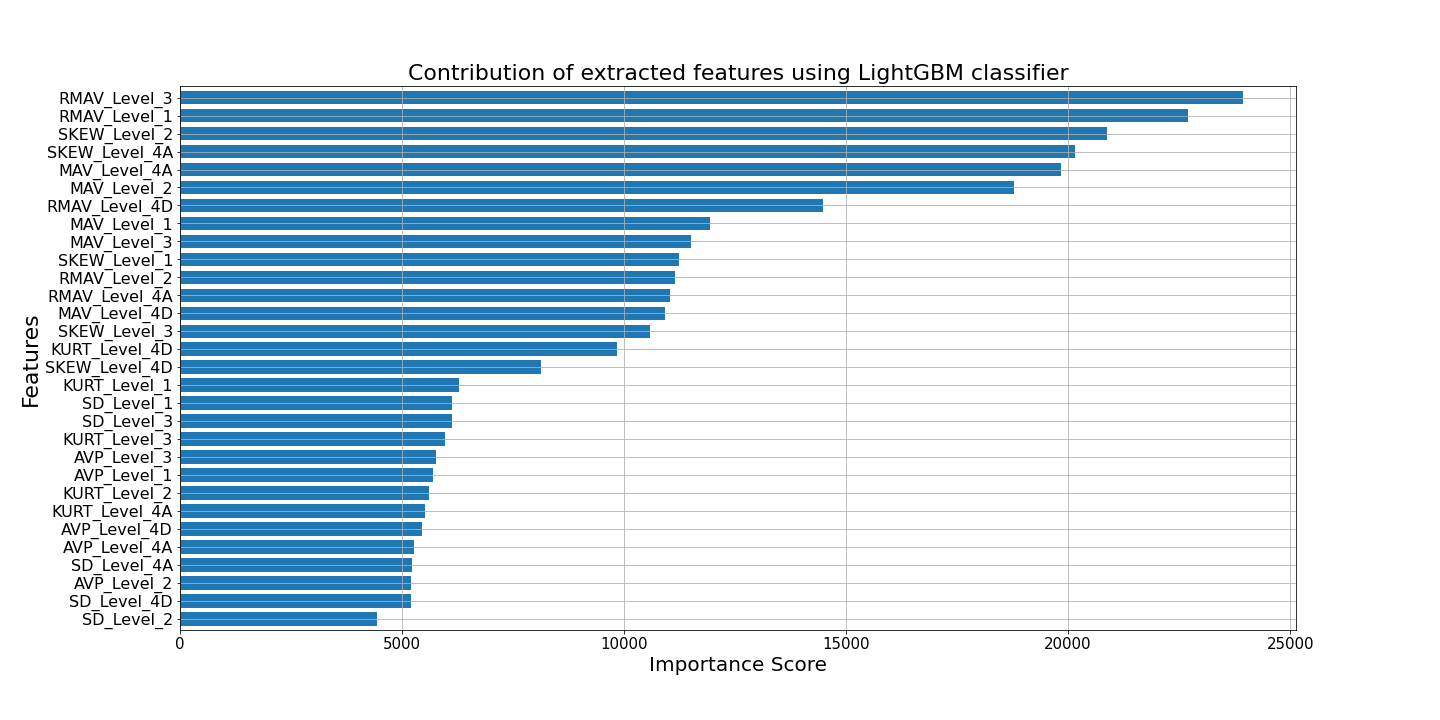}
    \caption{Importance scores for each feature obtained using LightGBM classifier. }
    \label{fig:lgbm_importance_score}
\end{figure*}

\textcolor{black}{We analysed the involved features to understand the importance of the features extracted by DTCWT. We used two feature analysis methods: filtering using ANOVA (Analysis of Variance) and LightGBM feature importance scores. In both techniques, selecting the top important features such as, 5,10 or 20 top features, always led to a decrease in classification results regardless of the choice of the number of selected features. \Cref{fig:anova} and \Cref{fig:lgbm_importance_score} present the results of features obtained by DTCWT using ANOVA and LightGBM feature importance respectively. We analysed the extracted features channel-wise and presented the average of all those channels’ features. 
It can be observed from the presented results that all extracted features by DTCWT in our technique play  important roles in improving the performance of classification. Therefore, we used all the features as we believe all features contribute to improve the classification results.}

\subsubsection{Classification}
\textcolor{black}{
As mentioned in \cref{INTRO}, we defined our problem in two classification problems: (1) Classification of seven seizure types, including both specific and non-specific seizure types in TUSZ (see \Cref{table:seiz_definition.0_1}), and (2) Classification of five seizure types, including only specific seizure types (see \cref{table:seiz_definition.0_1}).} 

\textcolor{black}{For both problems, we used Light Gradient Boosting Machine (LightGBM ver. 3.2.1) for classification.} LightGBM is a gradient boosting decision trees framework which utilizes a tree based learning algorithm. It is a proven to be an optimal choice to handle large size of data as it is memory efficient, trains faster and provides high accuracy~\cite{lightGBM}. The key characteristic of LightGBM is that it uses Gradient-based One-side Sampling (GOSS) in order to find the best split value. In addition, the exclusive feature bundling (EFB) technique is used in LightGBM to reduce the feature space complexity and the tree growth in LighGBM is leaf-wise growth that leads to faster training~\cite{lightGBM}. 
\textcolor {black}{
In a recent study of EEG binary classification for abnormality detection~\cite{albaqami2021automatic}, different classifiers were tested and LightGBM was found to be one of the most effective classifier in terms of results and training speed~\cite{lightGBM}.} Therefore, we selected LightGBM for classification. Hyperopt~\cite{bergstra2015hyperopt} was used to discover the best hyperparameters for our LightGBM.

\section{Results}\label{RESULTS}
In this paper, we used TUSZ EEG Corpus ver.1.5.2 to test our proposed technique for seizure type classification. 
Firstly, we applied some preprocessing methods to remove noise and to accentuate spike activity. Afterwards, DTCWT feature extraction method is applied and finally, LightGBM machine learning method is used for classification.

\subsection{Experimental settings}
\textcolor{black}{
A desktop computer with 16 GB main memory (RAM), 255 GB solid-state disk (SDD), 3.6 GHz microprocessor (CPU) and Windows 10 operating system is used for the experiments. The technique is developed in Python 3.7. using DTCWT Python package library ver. 0.12.0.}
\subsection{Performance evaluation}\label{Performance_evaluation}
\textcolor{black}{It can be observed from \cref{table:data_stats_v1.4.0} that TUSZ multi-class dataset suffers from the problem of class imbalance and the class distribution varies significantly. FNSZ, GNSZ and CPSZ classes have higher number of instances in the data as compared to the remaining classes. With this uneven class distribution, the accuracy alone cannot represent the performance of the proposed technique. Therefore, the average weighted F1-score is used to evaluate the performance of our proposed technique. Indeed, we report the average weighted Sensitivity, Specificity and Cohen's Kappa scores.}

As mentioned earlier, we applied our technique in two different classification problems: 7-class and 5-class classification. Moreover, we also tested our technique for both seizure-wise and patient-wise cross-validation classification. In seizure-wise cross-validation, we used a stratified 5-fold cross-validation which is inspired by state-of-the-art technique~\cite{Liu2020} in which the proportional distribution of classes in the entire dataset is randomly allocated to five-folds. This will also ensure a fair performance comparison with existing state-of-the-art research studies. For patient-wise cross-validation, we adopted the validation technique of Asif et al.~\cite{Asif} in which they applied 3-fold cross-validation across patients. In this scenario, the data presented in \cref{table:data_stats_v1.4.0} is split into three-folds. The selected classes of seizures include data from minimum of three patients. Therefore, this ensures that data used for testing is always from distinct patients whose data has never been used in the training phase.

\subsection{Experimental results}

In this section, we compare the obtained results for both evaluation scenarios. We present our proposed technique’s performance for 7-class problem followed by 5-class problem for each seizure-wise and patient-wise validation.

\subsubsection{Seizure-level cross-validation}
    
\begin{figure*}[!h] 
    \centering
    \includegraphics[type=pdf,ext=.pdf,read=.pdf,width=\textwidth]{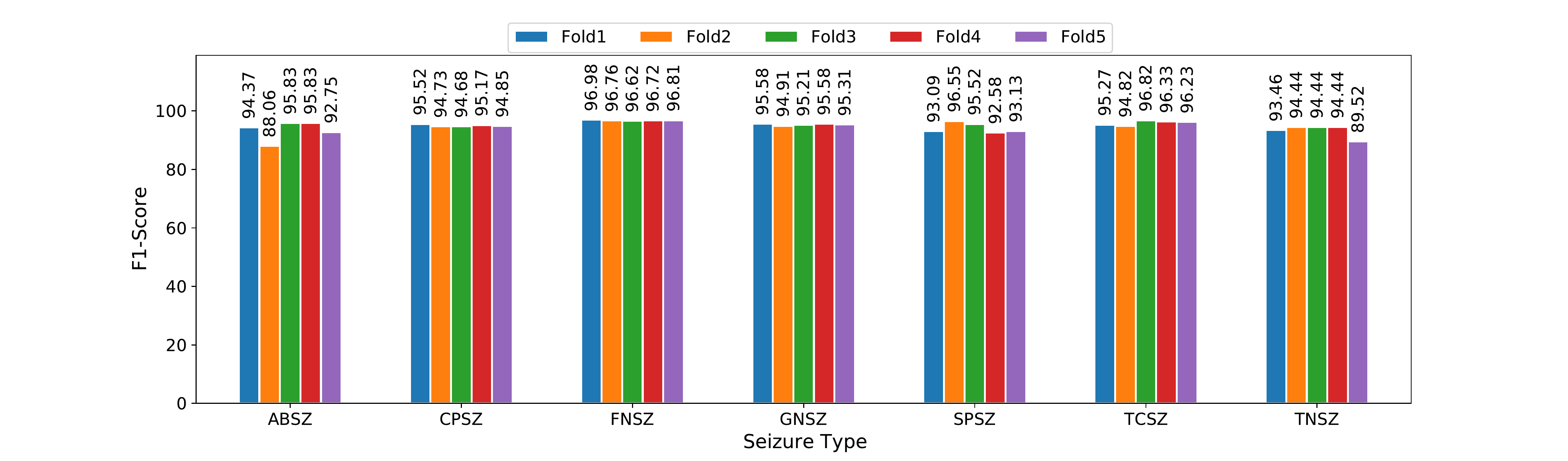}
    \caption{Performance of proposed technique on 7-class classification problem for each class having 5-fold cross-validation.}
    \label{fig:classififcation_results_of_differernt_tech}
\end{figure*}

        For both classification problems, we performed a 5-fold cross-validation. 
        For a 7-class problem, our proposed technique achieved the weighted average F1-score of 96.04\%.
        ~\Cref{fig:classififcation_results_of_differernt_tech} presents the classification performance in terms of F1-score for each class in the dataset for all 5-folds while \Cref{fig:figure3_7classes_sw} presents the confusion matrix for our proposed technique's performance on 7-class classification problem. 
        
\begin{figure*}[!h] 
    \centering
    \includegraphics[type=pdf,ext=.pdf,read=.pdf,width=\textwidth]{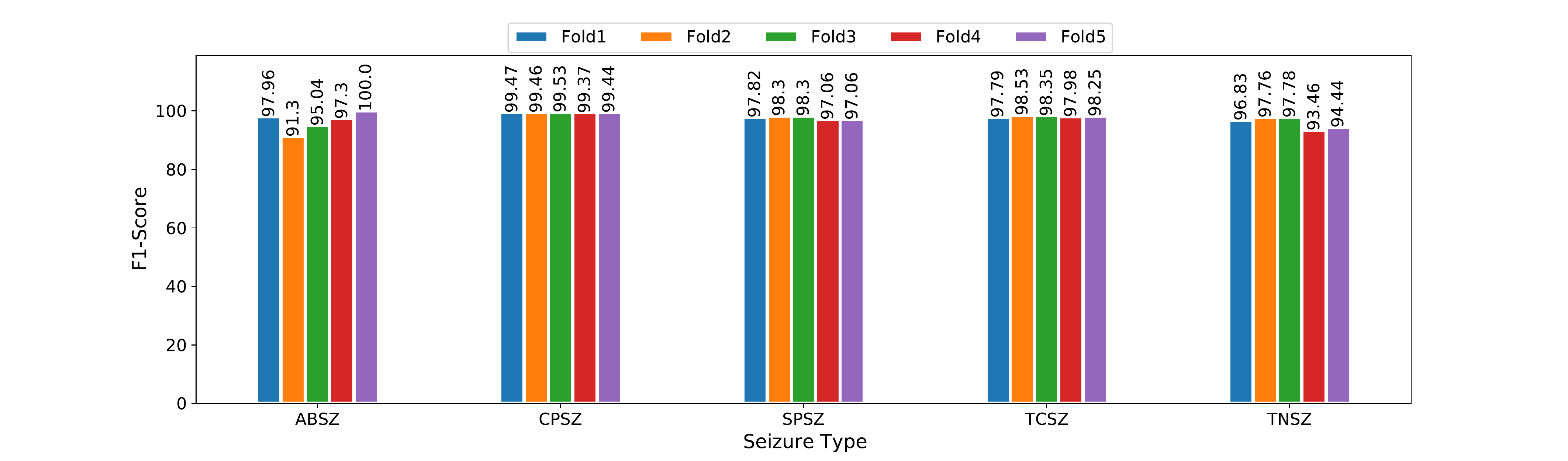}
    \caption{Performance of proposed technique on 5-class classification problem for each class having 5-fold cross-validation.}
    \label{fig:classififcation_results_5_classes}
\end{figure*}

\begin{figure*}[!h]
\centering
 \includegraphics[width=.9\linewidth]{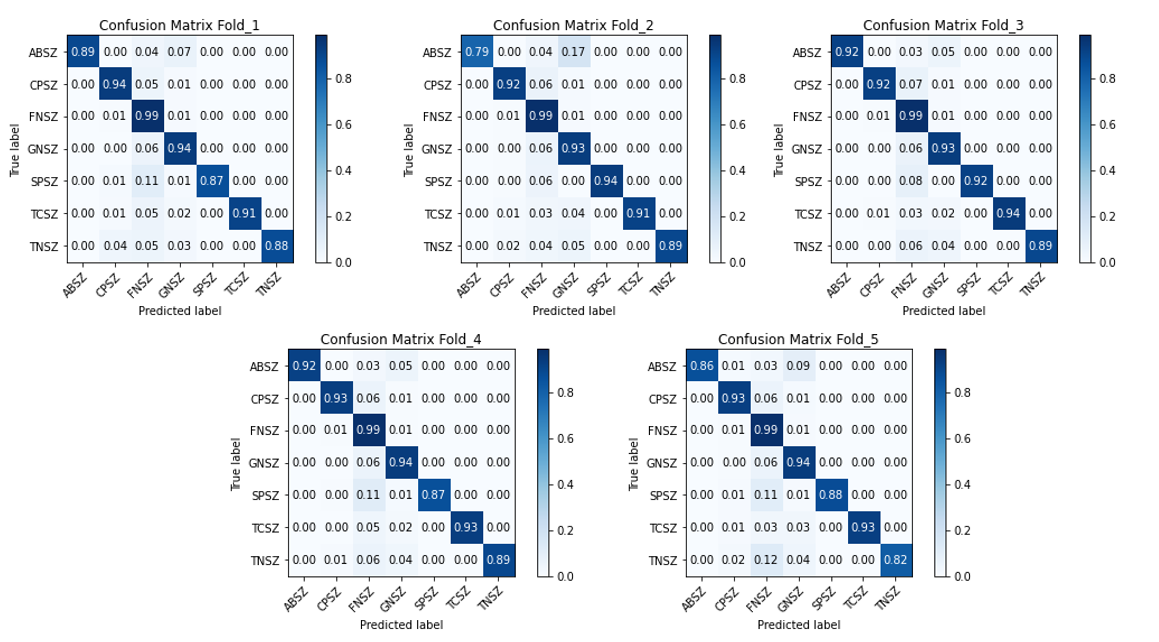}
\caption{Confusion matrix for 7-class classification problem having 5-fold cross-validation: 1st to 5th fold (Left to right, top to bottom).}
\label{fig:figure3_7classes_sw}

\end{figure*}

\begin{figure*}[!h]
\centering
 \includegraphics[width=.9\linewidth]{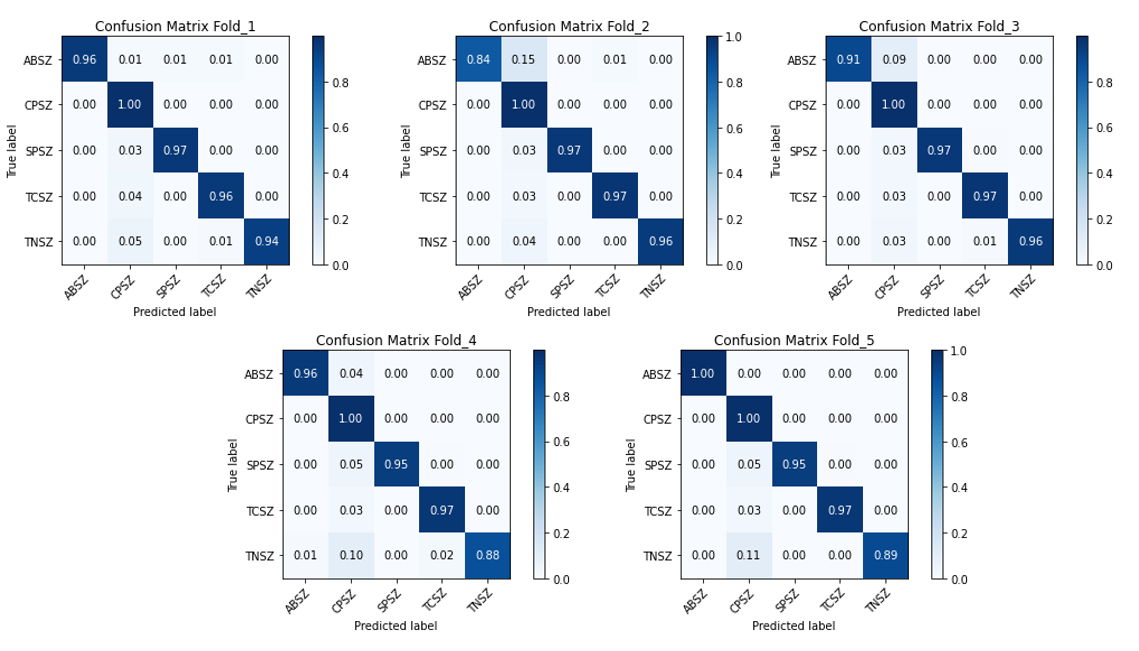}
\caption{Confusion matrix for 5-class classification problem having 5-fold cross-validation: 1st to 5th fold (Left to right, top to bottom).}
\label{fig:figure3_5classes_sw}

\end{figure*}

        For 5-class problem, when only the specific types of seizures are included (see \cref{table:data_stats_v1.4.0}), our method achieved weighted average F1-score of 99.1\%. This means that the non-specific seizures in the dataset have big impact on the performance of the machine learning algorithm as the results improved by more than 2\%; we discuss this in more detail in later sections. \Cref{fig:classififcation_results_5_classes} and \Cref{fig:figure3_5classes_sw} present the classification performance in terms of F1-score for each class in the dataset for all 5-folds and confusion matrix for 5-class classification problem respectively. 
        
        Moreover, the performance results of the proposed technique in terms of F1-score, Sensitivity, Specificity and Cohen's Kappa for each fold and for both classification problems are presented in~\Cref{table:table4newkappa}.

\subsubsection{Patient-wise cross-validation}

     For patient-wise cross-validation, 3-fold cross-validation was performed. We first evaluated our method for a 7-class problem and our proposed technique achieved the weighted average F1-score of 56.22\%. Similarly, for 5-class classification problem, the performance of our proposed technique significantly improved and the proposed technique achieved 75.97\% weighted average F1-score.

\begin{table}[hbt]
\scriptsize
\caption{Weighted average specificity, sensitivity, Cohen's Kappa and F1 score of the proposed method for 7-class and 5-class problems, having five-folds each}
  \centering
  \resizebox{\columnwidth}{!}{%
  \begin{tabular}{@{}cllccccc@{}}
  
    \toprule
                          &&  & {Specificity(\%)} & {Sensitivity(\%)} & {Cohen's Kappa(\%)}&F1 Score(\%)& 
                            \\\midrule
                            
   &\multirow{6}{*}{\rotatebox[origin=c]{90}{{{7-class problem}}}}

 &Fold 1             & 96.7&96.3&93.9&96.3&
                             \\
  & & Fold 2            & 96.4&95.9&93.3&95.9&
                             \\
   & &Fold 3             & 96.3&95.9&93.3&95.9&
     \\
    & &Fold 4             & 96.4&96.1&93.6&96.1&
     \\
    & &Fold 5          & 96.5&96.0&93.5&96.0&
     \\
     \cmidrule{3-7}
    & &Average             & 96.5&96.0&93.5&96.0&
     \\

         \midrule
         \midrule
                            
   &\multirow{6}{*}{\rotatebox[origin=c]{90}{{{5-class problem}}}}
   


   
 &Fold 1            & 97.2&99.1&97.5&99.1&
                             \\
  & & Fold 2            & 96.8&99.1&97.5&99.1&
                             \\
   & &Fold 3             & 97.4&99.2&97.8&99.2&
     \\
    & &Fold 4             & 96.6&98.9&97.0&98.9&
     \\
    & &Fold 5          & 96.8&99.1&97.4&99.1&
     \\
      \cmidrule{3-7}
    & &Average             & 97.0&99.1&97.4&99.1&
     \\

    \bottomrule
  \end{tabular}}
  \label{table:table4newkappa}
\end{table}

\section{Discussion}\label{discussion}
\cref{table:seizure-wise-all-studies} presents state-of-the-art techniques applied to the problem of seizure-type classification. It is difficult to compare the performance of our proposed technique with the existing studies in the literature, as each of the studies chose a different number of seizure classes. Therefore, we selected all the state-of-the-art studies considering more than 3 classes and compared our technique's performance with them as presented in~\cref{table:table_comparision_ptw}. 

It can be observed from~\cref{table:table_comparision_ptw} that our proposed technique's performance for specific seizure types classification is best among all the techniques at both seizure-level classification as well as patient-level classification, achieving F1-score of 99.1\% and 74.7\% respectively.~
For the 7-class problem, \cite{Asif} reported F1-score of 96.0\% using an ensemble architecture of three DenseNets. Similarly, the results of F1-score of 94.5\% and 90.1\% were reported in \cite{AhmedtAristizabal2020} and \cite{Roy2019} respectively. Our proposed technique outperformed all existing studies considering the same seven classes of seizures. 

\textcolor{black}{
For an eight-class problem, both~\cite{raghu2020eeg} and \cite{sriraam2019convolutional} proposed CNN-based solutions and reported the accuracy of 84.06\% and 88.3\% respectively. Similarly, Liu et al.,~\cite{Liu2020} reported high result of F1-score to be 97.4\% obtained by a symmetric bi-linear deep learning model consisting of two feature extractor models CNN and LSTM. The study demonstrates a limitation in testing. The 1-second segments considered in the dataset have 50\% overlap, which always has the potential of data leaking, as mentioned by~\cite{Liu2020}. Most of the work mentioned in \cref{table:table_comparision_ptw} is based on Fast Fourier Transform (FFT)~\cite{Roy2019,Asif,AhmedtAristizabal2020}, which has high resolution in the frequency-domain but zero resolution in the time-domain, which is very essential for EEG signal processing~\cite{rajoub2020characterization}. The other approaches~\cite{Liu2020,Raghu2020conf,raghu2020eeg} were based on Short-Time Fourier Transform (STFT) which is the known solution to overcome the limitation of FFT. STFT analyses the frequency of the signals at a particular short time period to avoid losing temporal information. However, STFT cannot catch sharp signal events because of the use of fixed windows length and fixed basis function~\cite{rajoub2020characterization}. On the other hand, our proposed technique overcomes these shortcomings by providing a smooth representation of EEG signals. It enables generating detailed features that have strong correlations with the latent structure of seizure types in EEG signals. Additionally, our proposed method also demonstrated very high classification results compared to other classical machine learning techniques~\cite{Roy2019,Wijayanto2019,Ramadhani2019,Saric2020}.}
Moreover, all of the research studies mentioned in \cref{table:table_comparision_ptw} utilized an older version of the TUSZ, which is ver.1.4.0, whereas the number of seizure events in the current version is much larger as compared to the previous version as shown in~\cref{table:data_stats_v1.4.0}. With a more challenging new version of TUSZ ver.1.5.2 which contains 1000 additional seizure events, our proposed method achieves better results for 7-class classification problem as compared to existing technique.


Most of the research studies in the literature chose to evaluate their methods only at the seizure-level. Out of the eight studies presented in \cref{table:seizure-wise-all-studies}, only \cite{Roy2019,Asif} considered the generalization of their models over different patients or in other words, model is trained and evaluated on data from different patients. This ensures that the performance of the model is general and can adopt for different patients. The performance of both studies~\cite{Roy2019,Asif} sharply decrease when evaluated using the patient-wise cross-validation technique as shown in~\cref{table:table_comparision_ptw}. Comparatively, our proposed method, showed a competitive result for 7-class problem and it demonstrated more stable performance when evaluated across different patients for 5-class problem. Our proposed technique achieved F1-score of 56.22\% and 74.7\% for classification of 7-class and 5-class problem respectively.

\textcolor {black}{
Regarding~\cite{swami2016novel}, DTCWT was employed to extract features from EEGs to classify epileptic vs. non-epileptic patients; however in this study, we are using the DTCWT with different set of features for a more complex problem which is the identification of the seizures types including specific and nonspecific focal and generalized seizures. Moreover, we evaluated this technique seizure-wise and patient-wise on the most extensive available EEG dataset, TUSZ ver.1.5.2~\cite{dataset2018}, containing data from more than 300 patients. In~\cite{swami2016novel}, data used for evaluation was obtained from only 21 subjects. This study has explored the generalization of our technique to be evaluated across different patients for better generalization capability.}

\begin{table*}
\scriptsize
\caption{Patient wise cross validation and seizure wise cross validation results for studies found in the literature.}
\centering
\begin{threeparttable}

  \centering
  \begin{tabularx}{0.9\linewidth}{XlXX}
    \toprule
        Method     & No. of Seizure Types & Seizure\_Wise CV(\%)  & Patient\_Wise CV(\%)\\
        \midrule

        SVM~\cite{Wijayanto2019}     & 4 &95.00  \tiny{Acc}&---\\
        SVM~\cite{Ramadhani2019}     & 4\tnote{*} &95.14  \tiny{Acc}& ---\\
        SeizureNet \cite{Asif}     & 7&95.00  \tiny{F1}    &62 \tiny{F1}\\
        KNN \cite{Roy2019}     & 7& 90.1  \tiny{F1} &40.1  \tiny{F1}   \\
        XGBoost \cite{Roy2019}     & 7& 85.1  \tiny{F1}     &   54.2  \tiny{F1}   \\
        SGD \cite{Roy2019}     & 7& 80.7  \tiny{F1}     &   46.9  \tiny{F1}  \\
        CNN \cite{Roy2019}     & 7& 71.8  \tiny{F1}     &   52.5  \tiny{F1}  \\
        NMN \cite{AhmedtAristizabal2020}& 7& 94.5  \tiny{F1}    &  ---    \\
        Inceptionv3~\cite{raghu2020eeg} &8\tnote{*} & 88.3 \tiny{Acc} &---\\
        AlexNet~\cite{sriraam2019convolutional}   & 8\tnote{*} &84.06 \tiny{Acc}&---\\
        CNN+LSTM \cite{Liu2020}& 8& 97.40  \tiny{F1}    &  ---    \\

        \textbf{This Work}     & 5& \textbf{99.1}  \tiny{F1}    &   \textbf{74.7}  \tiny{F1} \\
        \textbf{This Work}     & 7& 96.04  \tiny{F1}   &   56.22  \tiny{F1}  \\

    \bottomrule
  \end{tabularx}
  \label{table:table_comparision_ptw}
  \begin{tablenotes}                                      
    \item[*Including non-seizure EEG class.]  
    \end{tablenotes}
\end{threeparttable}

\end{table*}

In addition to the evaluation, we speculated that when only considering the EEG signals, it is not appropriate to treat the main seizures categories as unique seizure types against any of its sub-types. Since both SPSZ and CPSZ are sub-categories of focal seizures, it is unreasonable to train the machine learning algorithm to differentiate between them. According to the dataset, the reason for labeling an event as a focal non-specific seizure is the lack of information to make the decision~\cite{dataset2018}. After excluding the non-specific seizures labels, the experimental results demonstrated the stability of the classifier across different patients as shown in~\cref{table:table_comparision_ptw}. The performance of our proposed technique for specific-seizures type classification showed nearly perfect results when evaluated at the seizure level and it has the ability to generalize itself better on signals recorded from new patients as compared to~\cite{Roy2019,Asif}. Therefore, considering the non-specific seizures labels in TUSZ as unique types of seizure, when only utilizing EEG data, does not reveal meaningful results. Instead, one must include clinical features that neurologists look for when making a diagnosis (i.e., video EEG monitoring). By doing so, the machine learning algorithm knows the reason for labeling an event as non-specific, as there is not enough information to make the decision. This is beyond the scope of this paper as we focused solely on utilizing the EEG data. By knowing that the other specific seizure types in the corpus medically must be either focal or generalized at some point, we excluded the non-specific labels from our experiment and, in turn, the results demonstrated high and stable performance.

During patient-wise cross-validation, we noticed that the majority of the seizure events of type SPSZ classified as CPSZ. Again and from medical perspectives, we can relate this misclassification to the fact that the difference between focal CPSZ and focal SPSZ can majorly be determined by clinical characteristics, as described in~\cite{Fisher2017,dataset2018}. Therefore, considering the neurologist report in this situation might help in distinguishing between the two types. Moreover, as most of epileptic conditions are age-determined~\cite{sanei2013eeg}, we suggest that one could also include age, gender and medical history as an extra input features to the machine learning model to get more accurate results, which is beyond our scope in this paper.

\section{Conclusion}\label{conclusion}
Epilepsy is one of the most common neurological diseases that affect people of all ages. It is characterized by sudden and recurrent seizure attacks that appear in different forms and are treated in different ways. Correct assessment of epileptic seizures is vital in overcoming the complications of the disease, and it provides accurate information to the affected person. This paper presents a novel technique utilizing DTCWT and machine learning for automatic seizure type classification in EEGs. The proposed method demonstrates a significant improvement in classification, achieving 96.04\% and 99.1\% for seven-class and five-class classification problems respectively. We evaluated our proposed technique across different subjects, which is a very challenging task due to the limited amount of training data that is generalized to unseen test patients' EEG data. The achieved results show that our proposed technique perform significantly better as compared to the existing methods in the literature and is more general. The findings in this study enhances the applicability of artificial intelligence applications in assisting neurologists' decisions. In future research, we plan to investigate the use of different methods for feature extraction that can finely detect the differences between the seizures in EEG. 

\section*{Declaration of competing interest}
The authors have no conflicts of interest to declare.


\bibliographystyle{elsarticle-num-names}  

\end{document}